\documentclass[letterpaper, 10 pt, conference]{ieeeconf}  % Comment this line out if you need a4paper
\IEEEoverridecommandlockouts                              % This command is only needed if 
\usepackage{amsmath,amssymb,amsfonts}
\usepackage{graphicx}
\usepackage{algorithm}
\usepackage{algpseudocode}
\usepackage{todonotes}
\usepackage{multirow}
\usepackage{url}
\usepackage{cite}
\usepackage{textcomp}
\usepackage{xcolor}
\usepackage{optidef}

\usepackage{dblfloatfix}

\usepackage{tikz}
\usepackage{circuitikz}
\usetikzlibrary{arrows.meta}
\tikzstyle{block} = [draw, rectangle, 
    minimum height=3em, minimum width=6em]
\tikzstyle{sum} = [draw, circle, node distance=1cm]
\tikzstyle{input} = [coordinate]
\tikzstyle{output} = [coordinate]
\tikzstyle{pinstyle} = [pin edge={to-,thin,black}]

\DeclareMathOperator*{\argmin}{arg\,min}

\newcommand{\reals}{\mathbb{R}}
\newcommand{\BBM}{\begin{bmatrix}}
\newcommand{\EBM}{\end{bmatrix}}
\newcommand{\BEQ}{\begin{equation}}
\newcommand{\EEQ}{\end{equation}}
\DeclareMathOperator{\blkdiag}{blkdiag}

\title{\LARGE \bf Direct System Identification of Dynamical Networks with Partial Measurements: a Maximum Likelihood Approach}

\author{João Victor Galvão da Mata$^{1}$, Anders Hansson$^{2}$, and Martin S.~Andersen$^{1}$%<-this % stops a space
\thanks{*This work was supported by the Novo Nordisk Foundation under grant number NNF20OC0061894. It was also supported by ELLIIT.}
\thanks{$^{1}$Department of Applied Mathematics and Computer Science, Technical University of Denmark.
        Email: {\tt\small \{jogal,mskan\}@dtu.dk}}%
\thanks{$^{2}$Department of Electrical Engineering, Linköping University.
        Email: {\tt\small anders.g.hansson@liu.se}}%
}

\begin{document}

\maketitle
\thispagestyle{empty}
\pagestyle{empty}

\begin{abstract}
This paper introduces a novel direct approach to system identification of dynamic networks with missing data based on maximum likelihood estimation. Dynamic networks generally present a singular probability density function, which poses a challenge in the estimation of their parameters. By leveraging knowledge about the network's interconnections, we show that it is possible to transform the problem into a more tractable form by applying linear transformations. This results in a nonsingular probability density function, enabling the application of maximum likelihood estimation techniques. Our preliminary numerical results suggest that when combined with global optimization algorithms or a suitable initialization strategy, we are able to obtain a good estimate of the dynamics of the internal systems.
\end{abstract}

\begin{keywords}System identification, maximum likelihood estimation, dynamical networks, singular Gaussian distribution.
\end{keywords}

\section{Introduction}\label{sec:intro}
One of the greatest technical challenges in our society is to efficiently, sustainably and safely control large-scale complex systems.  These systems put new demands on control theory. Many of the available methods for modeling, analysis and design do not scale well with increasing complexity. Furthermore, the majority of control theory has been developed in a centralized setting, where all measurements are processed together to compute the control signals. While this paradigm offers conceptual advantages, it is not without inherent limitations. In reality, industrial practice frequently relies on distributed control structures, underscoring the need for more systematic approaches to design and analysis of such structures. The scope of applications spans a wide spectrum, encompassing networks for transportation, communication, and energy supply, as well as industrial production, logistics, and healthcare.

Design methodologies for controller design are often model based, and the time and effort needed for modeling is usually substantially larger than for controller design. Thus, achieving scalable control methods hinges on the ability to obtain these models in an efficient and scalable manner.

Large-scale complex systems are typically described as interconnections of simpler subsystems, i.e., as networked systems. We will here address system identification for such dynamic networks. These networks frequently contain internal variables that are not directly measurable, necessitating methods that are capable of handling what is termed ``latent variable'' in statistics and ``missing data'' within the system identification community. In essence, these methods should be able to perform identification tasks while relying on only partially observable variables.
%\cite{dankers2014system}.

When dealing with interconnected systems, there are essentially two ways to estimate the parameters of the systems: the \textit{direct approach} and the \textit{indirect approach}, which are explained for a single feedback loop in \cite{ljung1999book}. These approaches can be generalized to networks of systems. The direct approach is typically based on minimizing prediction errors to obtain the system parameters \cite{VANDENHOF20132994,ljung1999book}. The indirect approach first estimates parameters that characterize the closed loop transfer function between the inputs and the observable variables, and it then uses these functions and knowledge about the architecture of the system to estimate the parameters of the transfer function for the subsystems \cite{VANDENHOF19931523,hen+gev+baz19}.

As we will show with an example, the indirect approach suffers from the problem that even if all the true subsystems are stable and the closed loop is stable, the estimates of the subsystems may be unstable. It also suffers from the fact that the number of parameters that are needed to describe the closed loop system can be significantly larger than the total number of parameters describing the individual subsystems. As a result, the variance of the estimated models may be larger than if a direct method is used. However, current direct methods often require more variables to be measured in order for the method to give unbiased estimates using the prediction error method. We will show that using maximum likelihood (ML) estimation, we are able to obtain better estimates by just observing as many variables as used in the indirect method. 

Deriving the ML problem is challenging since networks of dynamical systems generally lead to a singular probability density function (pdf) and some variables are not observable. Before addressing this problem, we will first recapitulate the results of \cite{han+wal12} for ML estimation when not all variables are observable for a Gaussian distribution in Section~\ref{sec:nonsingluar}. Then, in Section~\ref{sec:sing-gaussian}, we will show how a singular Gaussian distribution can be transformed into a nonsingular one using linear transformations, and in Section~\ref{sec:dynamic-networks}, we apply this result to dynamic networks with known interconnections. We present a numerical example in Section~\ref{sec:num_exp}, and we conclude the paper in Section~\ref{sec:conclusions}.

\section{Nonsingular Gaussian Distribution and Maximum Likelihood Estimation}\label{sec:nonsingluar}

We start by considering a model of the form
\begin{align}\label{non_singular_gauss_distrib_case1}
    \Phi(\theta)x(\theta) + \Gamma(\theta) = e,
\end{align}
where $e \in \reals^m$ is a realization of a zero-mean Gaussian random variable with covariance $\lambda I$ with $\lambda > 0$, $\theta \in \reals^q$ is a vector of unknown model parameters, and $\Gamma(\theta) \in \mathbb{R}^{m}$. 

The vector
\begin{align*}
    x(\theta) = \begin{bmatrix}
        x_o(\theta) \\
        x_m(\theta)
    \end{bmatrix} \in  \mathbb{R}^{n}
\end{align*}
is composed of observed data $x_o\in\reals^{n_o}$ and missing data $x_m(\theta) \in \reals^{n_m}$, and we partition $\Phi(\theta) \in \reals^{m \times n}$ conformably as
$\Phi(\theta) = \begin{bmatrix}
    \Phi_o(\theta) & \Phi_m(\theta)
\end{bmatrix}$. To simplify notation, we will henceforth omit the dependence on $\theta$.

If $m=n$ and $\Phi$ is nonsingular, then the pdf for $x$ may be expressed as 
\begin{align*}
  p(x; \lambda,\theta) &= \mathcal N (x; \mu, \Psi)\\
  &= \frac{1}{\sqrt{(2\pi)^{m}\det{\Psi}}} \exp\left\{-\frac{1}{2\lambda}\|\Phi x+\Gamma\|_2^2\right\}
\end{align*}
where $\mu = -\Phi^{-1}\Gamma = (\mu_o, \mu_m)$ is the mean and $\Psi = \lambda(\Phi^T\Phi)^{-1}$ is the covariance matrix. Furthermore, the marginal pdf of $x_o$ may be expressed as 
\begin{align*}
    p(x_o; \lambda, \theta) = \mathcal N(x_o; \mu_o, \lambda (\Phi_o^T \Pi \Phi_o)^{-1})
\end{align*}
where $\Pi = I-\Phi_m (\Phi_m^T\Phi_m)^{-1}\Phi_m^T$ is a projection matrix.
As is shown in \cite{han+wal12}, up to an additive constant, the negative log-likelihood function can be expressed as
\begin{align}\label{eqn:cost_func}
\begin{split}
L(\lambda,\theta)=&\frac{1}{2\lambda}(\Phi_ox_o+\Gamma)^T\Pi (\Phi_ox_o+\Gamma)+
\frac{m}{2}\ln \lambda \\&-\frac{1}{2}\ln\det(\Phi^T\Phi)+\frac{1}{2}\ln\det Z
\end{split}
\end{align}
where $Z=\Phi_m^T\Phi_m$. An ML estimate of $(\lambda,\theta)$ can then be obtained as
\begin{align}\label{eqn:opt_prob}
(\hat \lambda, \hat \theta) \in \argmin_{(\lambda, \theta)} \, L(\lambda,\theta).  
\end{align}
We note that the partial derivatives of $L$ with respect to $\lambda$ and $\theta$ can be found in \cite{han+wal12}. 

\section{Singular Gaussian Distribution}\label{sec:sing-gaussian}

Dynamical networks generally lead to a singular pdf, making the derivation of the ML problem challenging. These interconnected systems, as we will show in Section~\ref{sec:dynamic-networks}, can be expressed as an instance of a more general model in the form of
\begin{align}\label{non_sing_gauss_distrib_1}
    A(\theta)x(\theta) + b(\theta) = \begin{bmatrix}
        e \\ 0
    \end{bmatrix},
\end{align}
where $e \in \mathbb{R}^{m_1}$ is a realization of a zero-mean Gaussian random variable with covariance $\lambda I$ with $\lambda >0$. As in the previous section, we partition $x(\theta)$ as 
\begin{align*}
    x(\theta) = \begin{bmatrix}
        x_o(\theta) \\
        x_m(\theta)
    \end{bmatrix} \in  \reals^{n_o + n_m},
\end{align*}
corresponding to observed and missing data, respectively, and we then partition $A(\theta) \in \reals^{(m_1+m_2) \times (n_o + n_m)}$ and $b(\theta) \in \reals^{m_1+m_2}$ conformably as
\begin{align*}
    A(\theta) = \begin{bmatrix}
    A_1(\theta) \\ A_2
\end{bmatrix} = \begin{bmatrix}
    A_{1o}(\theta)  & A_{1m}(\theta)\\  A_{2o}  & A_{2m}
\end{bmatrix}, \
b(\theta) = \begin{bmatrix}
    b_1(\theta) \\  b_2
\end{bmatrix}.
\end{align*}
Notice that $A_2$ and $b_2$ do not depend on the parameter vector $\theta$. We will see in Section~\ref{sec:dynamic-networks} that this assumption is satisfied for dynamic networks with known interconnections. Once again, we will omit the dependence on $\theta$ to simplify our notation.

In contrast to the situation in the previous section, the pdf for $x$ is now singular, and hence the maximum likelihood estimation is not readily applicable. We will address this issue by transforming the model under the assumption that $A(\theta)$ has full row-rank.

Given a singular value decomposition (SVD) of $A_{2m}$, i.e., 
\begin{align*}
    A_{2m} = \begin{bmatrix}
        U_1 & U_2
    \end{bmatrix}\begin{bmatrix}
    \Sigma_1 & 0 \\ 0 & 0    
    \end{bmatrix}\begin{bmatrix}
        V_1 & V_2
    \end{bmatrix}^T,
\end{align*}
we can rewrite (\ref{non_sing_gauss_distrib_1}) as
\begin{align*}
    \begin{bmatrix}
        A_{1o} & \Bar{A}_{1m1} & \Bar{A}_{1m2}\\ \Bar{A}_{2o1} & \Sigma_1 & 0 \\ \Bar{A}_{2o2} & 0 & 0
    \end{bmatrix}\begin{bmatrix}
        x_o \\ \Bar{x}_{m1} \\ \Bar{x}_{m2} 
    \end{bmatrix} + \begin{bmatrix}
        b_1 \\  \Bar{b}_{21} \\ \Bar{b}_{22} 
    \end{bmatrix} = \begin{bmatrix}
        e \\ 0 \\0
    \end{bmatrix}
\end{align*}
where
\begin{align}
\label{A2o2bar}
    \Bar{A}_{2o} &= \begin{bmatrix}
        \Bar{A}_{2o1} \\ \Bar{A}_{2o2} 
    \end{bmatrix} = \begin{bmatrix}
        U_1^T \\ U_2^T
    \end{bmatrix}A_{2o}, \\ 
    \label{A1m2bar}
    \Bar{A}_{1m} &= \begin{bmatrix}
        \Bar{A}_{1m1} & \Bar{A}_{1m2} 
    \end{bmatrix} = A_{1m}\begin{bmatrix}
        V_1 & V_2
    \end{bmatrix}
\end{align}
and
\begin{align*}
    \Bar{x}_{m} &= \begin{bmatrix}
        \Bar{x}_{m1} \\ \Bar{x}_{m2} 
    \end{bmatrix} = \begin{bmatrix}
        V_1^T \\ V_2^T
    \end{bmatrix}x_{m}, &
    \Bar{b}_{2} &= \begin{bmatrix}
        \Bar{b}_{21} \\ \Bar{b}_{22} 
    \end{bmatrix} = \begin{bmatrix}
        U_1^T \\ U_2^T
    \end{bmatrix}b_{2}.
\end{align*}
Using  $\Sigma_1$ as a pivot, we can rewrite the system as
\begin{align}
\label{eqn:manipuation1}
    \begin{bmatrix}
        \Bar{A}_{1o} & 0 & \Bar{A}_{1m2}\\ \Bar{A}_{2o1} & \Sigma_1 & 0 \\ \Bar{A}_{2o2} & 0 & 0
    \end{bmatrix}\begin{bmatrix}
        x_o \\ \Bar{x}_{m1} \\ \Bar{x}_{m2} 
    \end{bmatrix} + \begin{bmatrix}
        \Bar{b}_1 \\  \Bar{b}_{21} \\ \Bar{b}_{22} 
    \end{bmatrix} = \begin{bmatrix}
        e \\ 0 \\0
    \end{bmatrix},
\end{align}
where $\Bar{A}_{1o} = A_{1o} - \Bar{A}_{1m1} \Sigma_1^{-1}\Bar{A}_{2o1}$ and $\Bar{b}_1 = b_1 - \Bar{A}_{1m1} \Sigma_1^{-1}\Bar{b}_{21}$. The assumption that $A$ has full row-rank implies that $\Bar{A}_{2o2}$ has full row-rank, and hence there exists an orthogonal matrix $W = \begin{bmatrix} W_1 & W_2 \end{bmatrix}$ such that $\Bar{A}_{2o2}W = \begin{bmatrix}
    \Tilde{A}_{2o21} & 0
\end{bmatrix}$ with $\Tilde{A}_{2o21} = \Bar{A}_{2o2}W_1$ nonsingular. The matrix $W$ can be obtained by means of an SVD or an LQ decomposition of $\Bar{A}_{2o2}$. We then define
\begin{align}
\label{A1o2tilde}
    \begin{bmatrix}
        \Tilde{A}_{1o} \\
        \Tilde{A}_{2o1}
    \end{bmatrix} &= \begin{bmatrix}
        \Tilde{A}_{1o1} & \Tilde{A}_{1o2} \\
        \Tilde{A}_{2o11} & \Tilde{A}_{2o12}
    \end{bmatrix} = \begin{bmatrix}
        \Bar{A}_{1o} \\
        \Bar{A}_{2o1}
    \end{bmatrix}W, \\ \notag
    \Bar{x}_o &= \begin{bmatrix}
        \Bar{x}_{o1} \\
        \Bar{x}_{o2}
    \end{bmatrix} = W^Tx_o,
\end{align}
which allows us to rewrite equation \eqref{eqn:manipuation1} as
\begin{align*}
    \begin{bmatrix}
        \Tilde{A}_{1o1} & \Tilde{A}_{1o2}  & 0 & \Bar{A}_{1m2}\\ 
        \Tilde{A}_{2o11} & \Tilde{A}_{2o12} & \Sigma_1 & 0 \\
        \Tilde{A}_{2o21} & 0 & 0 & 0
    \end{bmatrix}\begin{bmatrix}
        \Bar{x}_{o1} \\
        \Bar{x}_{o2} \\
        \Bar{x}_{m1} \\ 
        \Bar{x}_{m2} 
    \end{bmatrix} + \begin{bmatrix}
        \Bar{b}_1 \\  \Bar{b}_{21} \\ \Bar{b}_{22} 
    \end{bmatrix} = \begin{bmatrix}
        e \\ 0 \\0
    \end{bmatrix}.
\end{align*}

\vspace*{-7px}
\noindent Using $\Tilde{A}_{2o21}$ as a pivot, we obtain the system
\begin{align*}
    \begin{bmatrix}
        0 & \Tilde{A}_{1o2}  & 0 & \Bar{A}_{1m2}\\ 
        \Tilde{A}_{2o11} & \Tilde{A}_{2o12} & \Sigma_1 & 0 \\
        \Tilde{A}_{2o21} & 0 & 0 & 0
    \end{bmatrix}\begin{bmatrix}
        \Bar{x}_{o1} \\
        \Bar{x}_{o2} \\
        \Bar{x}_{m1} \\ 
        \Bar{x}_{m2} 
    \end{bmatrix} + \begin{bmatrix}
        \Tilde{b}_1 \\  \Bar{b}_{21} \\ \Bar{b}_{22} 
    \end{bmatrix} = \begin{bmatrix}
        e \\ 0 \\0
    \end{bmatrix},
\end{align*}
where $\Tilde{b}_1 = \Bar{b}_1 - \Tilde{A}_{1o1}\Tilde{A}_{2o21}^{-1} \Bar{b}_{22}$. Finally, we rewrite this system as two systems,
\begin{align}
\label{equiv_non_sing}
    \begin{bmatrix}
        \Tilde{A}_{1o2} & \Bar{A}_{1m2}
    \end{bmatrix} \begin{bmatrix}
        \Bar{x}_{o2} \\
        \Bar{x}_{m2} 
    \end{bmatrix} + \Tilde{b}_1 = e, \\
    \begin{bmatrix}
        \Tilde{A}_{2o11} & \Sigma_1\\
        \Tilde{A}_{2o21} & 0
    \end{bmatrix}\begin{bmatrix}
        \Bar{x}_{o1} \\
        \Bar{x}_{m1}
    \end{bmatrix}+\begin{bmatrix}
        \Bar{b}_{21} + \Tilde{A}_{2o12}\Bar{x}_{o2}\\
        \Bar{b}_{22}
    \end{bmatrix} = \begin{bmatrix}
        0 \\ 0
    \end{bmatrix}.
\end{align}
From this we realize that only $\bar x_{02}$ and $\bar x_{m2}$ are directly related to $e$. The matrix $\BBM \tilde A_{1o2}&\bar A_{1m2}\EBM$ has full row-rank, so if it is a square matrix, then it is also invertible, and otherwise we can make use of column compression to further reduce the number of variables. 

The equation \eqref{equiv_non_sing} is of the form \eqref{non_singular_gauss_distrib_case1} with 
    \[ \Phi = \begin{bmatrix} \Tilde{A}_{1o2} & \Bar{A}_{1m2} \end{bmatrix}, \quad \Gamma = \Tilde{b}_1, \] 
and hence we can processed as for the nonsingular Gaussian distribution in Section~\ref{sec:nonsingluar}. We note that $\Phi$ is a linear transformation of $A_1$. To see this, first note that $\Tilde{A}_{1o2} =  \Bar{A}_{1o}W_2$, which follows from \eqref{A1o2tilde}, and recall that \[\Bar{A}_{1o} = A_{1o} - \Bar{A}_{1m1} \Sigma_1^{-1}\Bar{A}_{2o1} = A_{1o} - A_{1m}V_1 \Sigma_1^{-1}U_1^T A_{2o}.\] Using \eqref{A1m2bar}, we arrive at
\begin{align*}
    \Phi &= \begin{bmatrix}
        \Phi_o & \Phi_m
    \end{bmatrix} \\
    &= \begin{bmatrix}
    A_{1o} & A_{1m}
\end{bmatrix} \begin{bmatrix}
    W_2 & 0 \\ - V_1 \Sigma_1^{-1}U_1^T A_{2o} W_2 & V_2
\end{bmatrix}.
\end{align*}
Similarly, tracing the transformations applied to $b$, we find that
\begin{align*}
    \Gamma = b_1 - \begin{bmatrix}     A_{1o} & A_{1m}
    \end{bmatrix} \begin{bmatrix}
        H b_2 \\ V_1 \Sigma_1^{-1}U_1^T(I-A_{2o}H)b_2
    \end{bmatrix},
\end{align*}
where $H = W_1(U_2^TA_{2o}W_1)^{-1}U_2^T$.
Thus, the reduced nonsingular problem is obtained as a linear transformation of $(A_1, b_1)$ where the transformation matrices are functions of $(A_2, b_2)$, which are known and do not depend on $\theta$.

\section{Dynamic Networks}\label{sec:dynamic-networks}
We will now consider a dynamic network model that can be cast in the form of 
\eqref{non_sing_gauss_distrib_1} where $A_2$ and $b_2$ do not dependent on the parameters vector $\theta$. Specifically, will consider a network of $M$ systems, where the $i$th system is described by an ARMAX model of the form
\begin{align}
\label{eqn:ARMAX}
y_k^i= -\sum_{j=1}^{n_a^i} a_j^iy_{k-j}^i  + \sum_{j=0}^{n_b^i} b_j^i u_{k-j}^i + e_k^i + \sum_{j=1}^{n_c^i} c_j^i e_{k-j}^i
\end{align}
for $k=1,2,\ldots,N$. We will assume that $y_k^i$, $u_k^i$, and $e_k^i$ are zero for all $k\leq 0$. To simplify notation, we define $u^i=(u_1^i,\ldots,u_N^i)$, $y^i=(y_1^i,\ldots,y_N^i)$, and $e^i=(e_1^i,\ldots,e_N^i)$, corresponding to the $i$th system's input, output, and disturbance signals. We also define $a^i=(a_1^i,\ldots,a_{n_a^i}^i)$, $b^i=(b_0^i,\ldots,b_{n_b^i}^i)$, and $c^i=(c_1^i,\ldots,c_{n_c^i}^i)$ as well as lower-triangular Toeplitz matrices $T_{a^i}\in\reals^{N\times N}$, $T_{b^i}\in\reals^{N\times N}$ and $T_{c^i}\in\reals^{N\times N}$ whose first columns are
$$ \BBM 1\\a^i\\0\EBM,\quad\BBM b^i\\0\EBM,\quad\BBM 1\\c^i\\0\EBM,$$
respectively, c.f., \cite{wal+han14}. This allows us to express the ARMAX model \eqref{eqn:ARMAX} as
\begin{align*}
    T_{a^i} y^i=T_{b^i}u^i + T_{c^i} e^i.
\end{align*}
The interconnections are defined in terms of sparse matrices $\Lambda\in\reals^{M\times M}$ and 
$\Omega\in\reals^{M\times Q}$ with $\pm 1$ as nonzero entries such that
$$\BBM u_k^1\\u_k^2\\\vdots\\u_k^M\EBM=\Lambda \BBM y_k^1\\y_k^2\\\vdots\\y_k^M\EBM +
\Omega \BBM r_k^1\\r_k^2\\\vdots\\r_k^Q\EBM, \quad 1\leq k \leq N,$$
where $r^i=(r_1^i,r_2^i,\ldots,r_N^i)$, $i = 1,\ldots,Q$, are exogenous signals. We will assume that $r_k^i=0$ for all $k\leq 0$.

We will now show that the dynamic network model can be written as in \eqref{non_sing_gauss_distrib_1}. To this end, we  
define vectors $y=(y^1,\ldots,y^M)$, $u=(u^1,u^2,\ldots,u^M)$, 
$e=(e^1,e^2,\ldots,e^M)$, and $r=(r^1,r^2,\ldots,r^Q)$, and matrices
$T_{y^i}=T_{c^i}^{-1}T_{a^i}$ and $T_{u^i}=T_{c^i}^{-1}T_{b^i}$ for $i=1,\ldots,M$. We also define two block-diagonal matrices,
\begin{align*} 
    T_y &=  \blkdiag (T_{y^1},\ldots,T_{y^M}) \\
    T_u &= \blkdiag (T_{u^1},\ldots,T_{u^M}).
\end{align*}
This allows us to express the dynamic network model as an instance of equation \eqref{non_sing_gauss_distrib_1} with
\begin{align*} 
A&=\BBM A_1\\A_2\EBM = \BBM T_y & -T_u \\ -\Lambda\otimes I& I\EBM(P\otimes I) \\
x&=(P\otimes I)^T\BBM y\\u\EBM\\
b&=\BBM b_1\\b_2\EBM= -\BBM 0\\ \Omega\otimes I\EBM r,
\end{align*}
where $P \in \reals^{2M \times 2M}$ is a permutation matrix that is defined such that the observed parts of 
$y$ and $u$ correspond to the leading entries of $x$. The parameter vector $\theta$ represents the unknown model parameters $(a^1,\ldots,a^M,b^1,\ldots,b^M,c^1,\ldots,c^M)$.

The matrix $A$ satisfies $m_1=m_2=MN$ and $n = 2MN$, and hence it is square. Using the fact that the matrices $T_{a^i}$ and $T_{c^i}$ are nonsingular for all $i$, we see that $T_y$ is nonsingular and then, using the Schur complement, we have that $A$ is nonsingular if $T_y - T_u(\Lambda\otimes I)$ is full rank. As a result, the transformation described in Section~\ref{sec:sing-gaussian} yields a square and nonsingular matrix $\Phi=\BBM \bar A_{1o2}&\bar A_{1m2}\EBM$ as in
\eqref{equiv_non_sing}. Furthermore, $A_2$ only depends on the network topology and is independent of any model parameters. 
Thus, the linear transformations that are needed to obtain $\Phi$ and $\Gamma$ are independent of the 
model parameters.

\section{Numerical Experiment}\label{sec:num_exp}
\begin{figure}[htbp]
\begin{center}
\resizebox{0.48\textwidth}{!}{
\begin{tikzpicture}[auto, node distance=2cm]
    % We place the blocks, the inputs, and the summations
    \node [input, name=r2] {};
    \node [block, right of=r2, node distance=4cm] (G2) {$G^2$};
    \node [sum, right of=G2, node distance=4cm] (sum2) {$+$};
    \node [input, above of=sum2] (e2) {};
    \node [block, below of=G2, node distance=3cm] (G3) {$G^3$};
    \node [sum, right of=G3, node distance=2cm] (sum3) {$+$};
    \node [sum, right of=sum3, node distance=2cm] (sum23) {$+$};
    \node [input, above of=sum3] (e3) {};
    \node [sum, below of=sum23, node distance=3cm] (sum4) {$+$};
    \node [sum, left of=G3, node distance=2cm] (sum13) {$+$};
    \node [block, below of=G3, node distance=3cm] (G1) {$G^1$};
    \node [sum, left of=G1, node distance=2cm] (sum1) {$+$};
    \node [input, below of=sum1] (e1) {};
    \node [input, left of=sum13] (r3) {};
    \node [input, right of=sum4] (r1) {};
    % We draw edges
    \draw [-Latex] (r1) -- node {$r^1$} (sum4);        
    \draw [-Latex] (r2) -- node {$r^2=u^2$} (G2);    
    \draw [-Latex] (r3) -- node {$r^3$} (sum13);
    \draw [-Latex] (e1) -- node {$e^1$} (sum1);
    \draw [-Latex] (e2) -- node {$e^2$} (sum2);
    \draw [-Latex] (e3) -- node {$e^3$} (sum3);
    \draw [-Latex] (G2) -- node {} (sum2);
    \draw [-Latex] (sum2) -- node {$y^2$} (sum23);
    \draw [-Latex] (G3) -- node {} (sum3);
    \draw [-Latex] (sum3) -- node {$y^3$} (sum23);
    \draw [-Latex] (sum23) -- node {} (sum4);
    \draw [-Latex] (sum4) -- node {$u^1$} (G1);
    \draw [-Latex] (G1) -- node {} (sum1);
    \draw [-Latex] (sum1) -- node {$y^1$} (sum13);
    \draw [-Latex] (sum13) -- node {$u^3$} (G3);
\end{tikzpicture}
}
\end{center}
\caption{Block diagram for a dynamic network.}
\centering
\label{fig:dyn-network}
\end{figure}
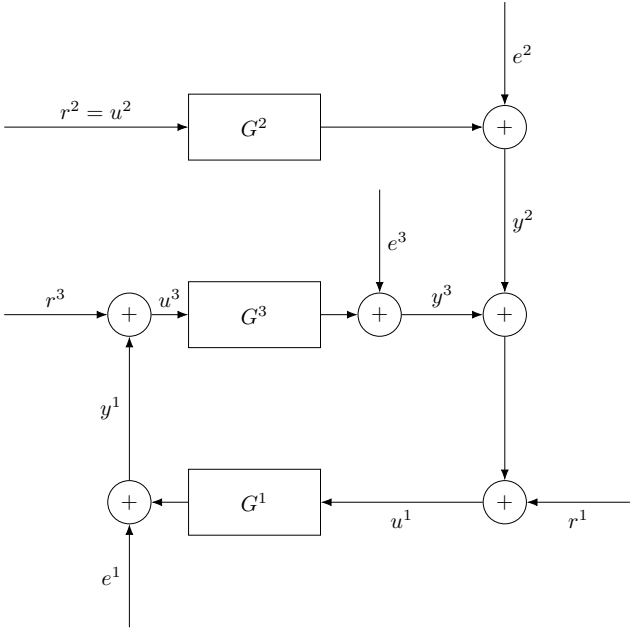

We will now illustrate some properties of the proposed method using a numerical example based on the network of $M=3$ systems shown in Fig.~\ref{fig:dyn-network}. The interconnections are described by the matrices
$$\Lambda=\BBM 0&1&1\\0&0&0\\1&0&0\EBM, \quad \Omega=I.$$
The example matches the example in Fig.~2 in \cite{hen+gev+baz19} except for the way the disturbances enter the system. 

Each subsystem is a second-order system, and we only have one observable variable, namely $x_o = (u^3)$. As we will show soon, the indirect approach does not handle ARMAX models, so we will limit our attention to the case where all subsystems are ARX models in order to facilitate a fair comparison. We generated $N=500$ measurements where the error $(e^1,e^2,e^3)$ is a realization of a zero-mean Gaussian random variable with covariance $0.1\cdot I$, and the input $(r^1,r^2,r^3)$ is a vector of independent samples from the Rademacher distribution. 

The true systems that we use to generate both the observable and the missing states are zero-order hold discretizations of continuous systems whose Laplace domain transfer functions are given by
\begin{align*}
G^1 = \frac{0.5}{(s+2)^2},\,\,\,
G^2 = \frac{1}{s^2+2s+0.75},\,\,\,
G^3 = \frac{0.5}{(s+0.5)^2}.
\end{align*}
The three systems are stable, and hence the poles of the discrete models will be inside the unit circle.

\subsection{Direct Approach}\label{ssec:direct_approach}

We will use gradient descent combined with a backtracking line search to find local minima of the negative log-likelihood function, which is not a convex function. Given the observed data $x_o$, we generate 100 initial parameter vectors $\theta$ such that the closed-loop system is stable. The parameters for each system are drawn from a uniform distribution on $[-1,1]$. 

In order to reduce the complexity of the optimization problem \eqref{eqn:opt_prob} we rewrote the cost function to eliminate the dependence on $\lambda$. Minimizing \eqref{eqn:cost_func} with respect to $\lambda$ leads to
\begin{align*}
    \lambda^\star = \frac{1}{m} (\Phi_ox_o+\Gamma)^T\Pi (\Phi_ox_o+\Gamma),
\end{align*}
and then the cost function can be rewritten as
\begin{align*}
\begin{split}
L(\theta)=\, &\frac{m}{2}\ln(\frac{1}{m}(\Phi_ox_o+\Gamma)^T\Pi (\Phi_ox_o+\Gamma)) \\&-\frac{1}{2}\ln\det(\Phi^T\Phi)+\frac{1}{2}\ln\det Z+\frac{m}{2}.
\end{split}
\end{align*}
Our experiments showed that performing this variable reduction leads to better convergence properties than solving the original problem.

Our implementation is based on the Python library JAX \cite{jax}, which uses automatic differentiation to compute the partial derivatives of the cost function.

\subsection{Indirect Approach}\label{ssec:indirec_approach}
The indirect approach first estimates the overall transfer function between the inputs and the observable variables, and then it uses this information to estimate a transfer function for each subsystem separately.

To analyze the example using transfer functions, we first write the ARMAX model in \eqref{eqn:ARMAX} as
$$y_k^i=G^i(q)u_k+H^i(q)e_k^i$$
where $q$ denotes the forward shift operator such that $qu_k=u_{k+1}$, and where
$$G^i(q)=\frac{B^i(q)}{A^i(q)},\qquad 
H^i(q)=\frac{C^i(q)}{A^i(q)}$$
and
\begin{align*}
A^i(q)&=q^{n_a^i}+a_1^iq^{n_a^i-1} +\cdots+ a_{n_a^i}^i\\
B^i(q)&=b_0^iq^{n_b^i}+b_1^iq^{n_b^i-1} +\cdots+ b_{n_b^i}^i\\
C^i(q)&=c_0^iq^{n_c^i}+c_1^iq^{n_c^i-1} +\cdots+ c_{n_c^i}^i.
\end{align*}
It is straight forward to show using algebraic manipulations that
\begin{align*}
    \Delta(q)u^3_k=&G^1(q)r^1_k+G^1(q)G^2(q)r^2_k+r^3_k+
    H^1(q)e^1_k\\&+G^1(q)H^2(q)e^2_k+G^1(q)H^3(q)e^3_k
\end{align*}
where $\Delta(q)=1-G^1(q)G^3(q)$. Equivalently, we can write
\begin{align*}
    A^2(A^1A^3-B^1B^3)u^3_k=&B^1A^2A^3r^1_k+B^1B^2A^3r^2_k\\ &+A^1A^2A^3r^3_k
    +C^1A^2A^3e^1_k\\ &+B^1C^2A^3e^3_k+B^1A^2C^3e^3_k
\end{align*}
where we have omitted the dependence on $q$ to simplify notation. 
From this we realize that it is possible to use an ARMAX model of the form
\begin{align}\label{eqn:armax_model_indirect}
    \bar Au^3_k=\bar B^1r_k^1+\bar B^2r_k^2+\bar B^3r_k^3+
\bar C\bar e_k
\end{align}
where we only have measurements of $u_k^3$ and $(r_k^1,r_k^2,r_k^3)$ in order to estimate 
the model using any traditional identification technique. In our numerical experiments, we used the System Identification Toolbox \cite{ljung1995system} from MATLAB. It then follows that 
\begin{align*}
\frac{B^1}{A^1}&=\frac{\bar B^1}{\bar B^3}\\
\frac{B^2}{A^2}&=\frac{\bar B^2}{\bar B^1}\\
\frac{B^3}{A^3}&=\frac{\bar B^3-\bar A}{\bar B^1}
\end{align*}
from which $A^i$ and $B^i$ can be found. Notice that it is not possible to recover $C^i$. Thus, we will from now on assume 
that we have ARX models, i.e., $C^i=1$.

If we assume that all the internal systems are order two, the polynomials $A^i(q)$ and $B^i(q)$ will have degree two, then the number
of parameters are 5 for each ARX model, i.e., a total of 15 parameters. 
However, we have that $\bar A$ and $\bar B^i$ are of degree 6
whereas $\bar C$ is of degree
of degree 4. We will then need 26 parameters when estimating the reformulated model described in \eqref{eqn:armax_model_indirect}. To prevent an increase in the number of parameters, one can either add constraints to the problem or perform a reparametrization of the cost function. Opting for this approach comes with the trade-off of increasing the computational complexity.

Here we have normalized such that $\bar A$ and $\bar C$ are monic polynomials.

For this example the number of parameters have increased using the 
indirect approach. This may impair the quality of the estimate, since it is known
that the more parameters that are estimated, the higher is the variance of the 
estimate. However, the indirect approach benefits from that a standard prediction
error method may be used to obtain unbiased estimates, and these estimates
may be utilized to initialize the algorithm. Moreover, the analysis shows 
what signals need to be measured and which do not need to be measured. 
Here we see that it is enough to measure $u^3$ in addition to $r$. 
Similar analysis can be carried out to show that either of 
$u^1$, $u^3$, $y^1$ or $y^3$ are sufficient to measure, c.f. the example 
related to Fig.~2 in \cite{hen+gev+baz19}. This reference also discusses what 
signals need to be measured for a general dynamic network in order to 
estimate all parameters when using indirect approaches.

\subsection{Results}\label{ssec:results}
When observing only $x_o = u^3$, we found that all 100 different initial points generated for the direct approach converged to local minima corresponding to stable systems. For most of these estimates, the missing states $x_m$ were poorly estimated, while the observed states $x_o$ were well estimated. Only the best local solutions (in terms of the cost function value) led to good estimates of both the observable and missing states. 

Using the indirect approach, $\bar B^1$ presented unstable zeros, which would result in unstable models for the subsystems 2 and 3. To mitigate this, the unstable zeros where removed from $\bar B^1$ and the static gain was adjusted in order to match its original value before the exclusion of the unstable zeros.

\begin{figure}[hbt!]
    \centering
\includegraphics[width=0.5\textwidth]{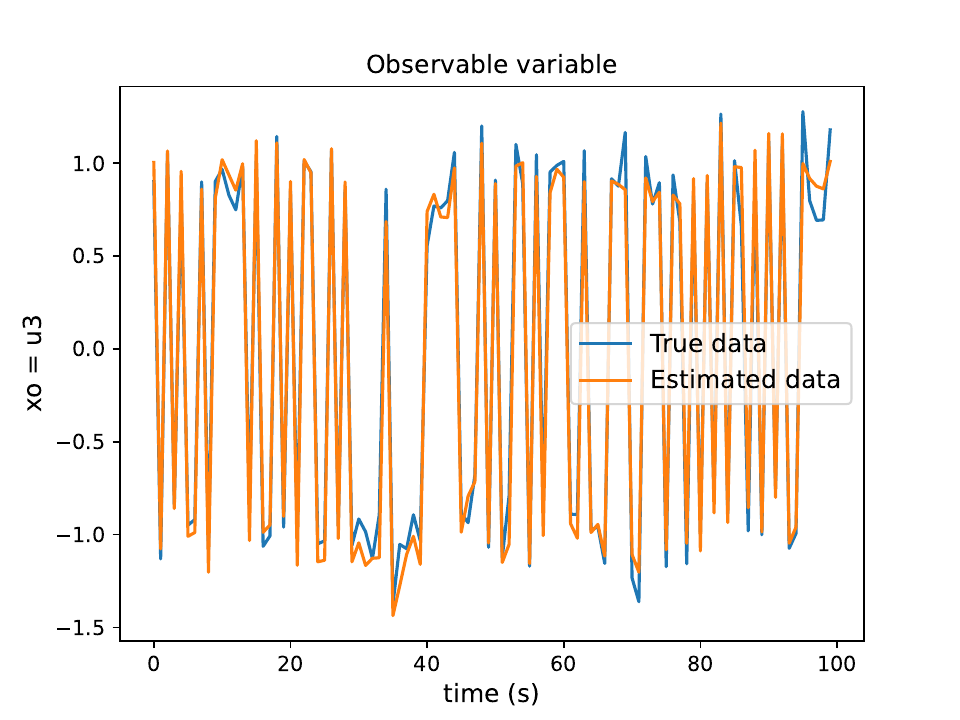}
\caption{Estimated and true observable variables.} \label{observable_results}
\end{figure}

\begin{figure*}[htbp]
\vspace{3px}
    \centering
\includegraphics[width=\textwidth]{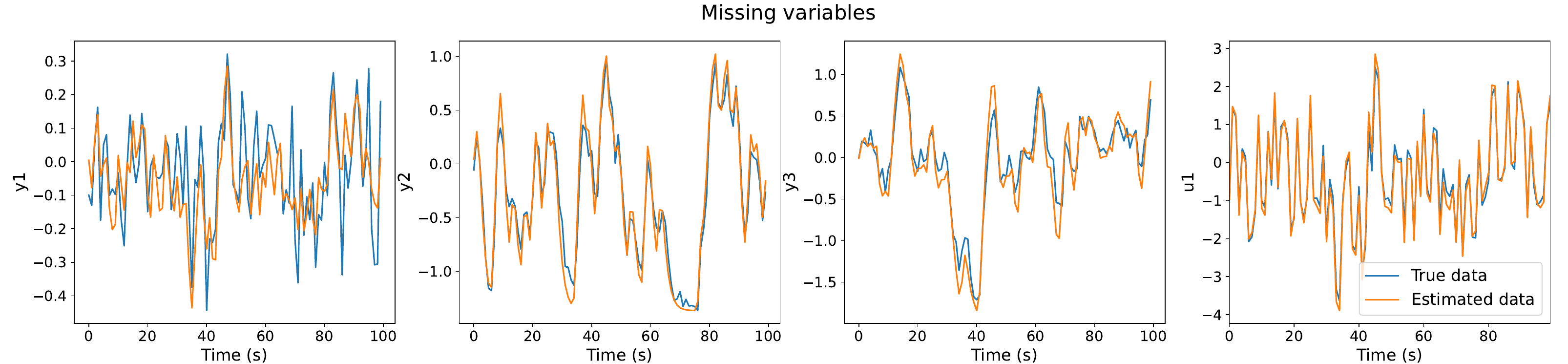}
\caption{Comparison between the true missing variables and the estimate obtained using the proposed direct approach.} \label{missing_results}
\end{figure*}

Fig.~\ref{observable_results} presents the comparison between the true observed variable ($u_3$) and the one generated when we simulated the system using the estimated parameter vector $\hat \theta$ from the proposed direct approach with the lowest cost function value. Fig.~\ref{missing_results} features a similar comparison for the missing variables ($y_1,y_2,y_3,u_1$).

\begin{figure*}[htbp]
    \centering
\includegraphics[width=\textwidth]{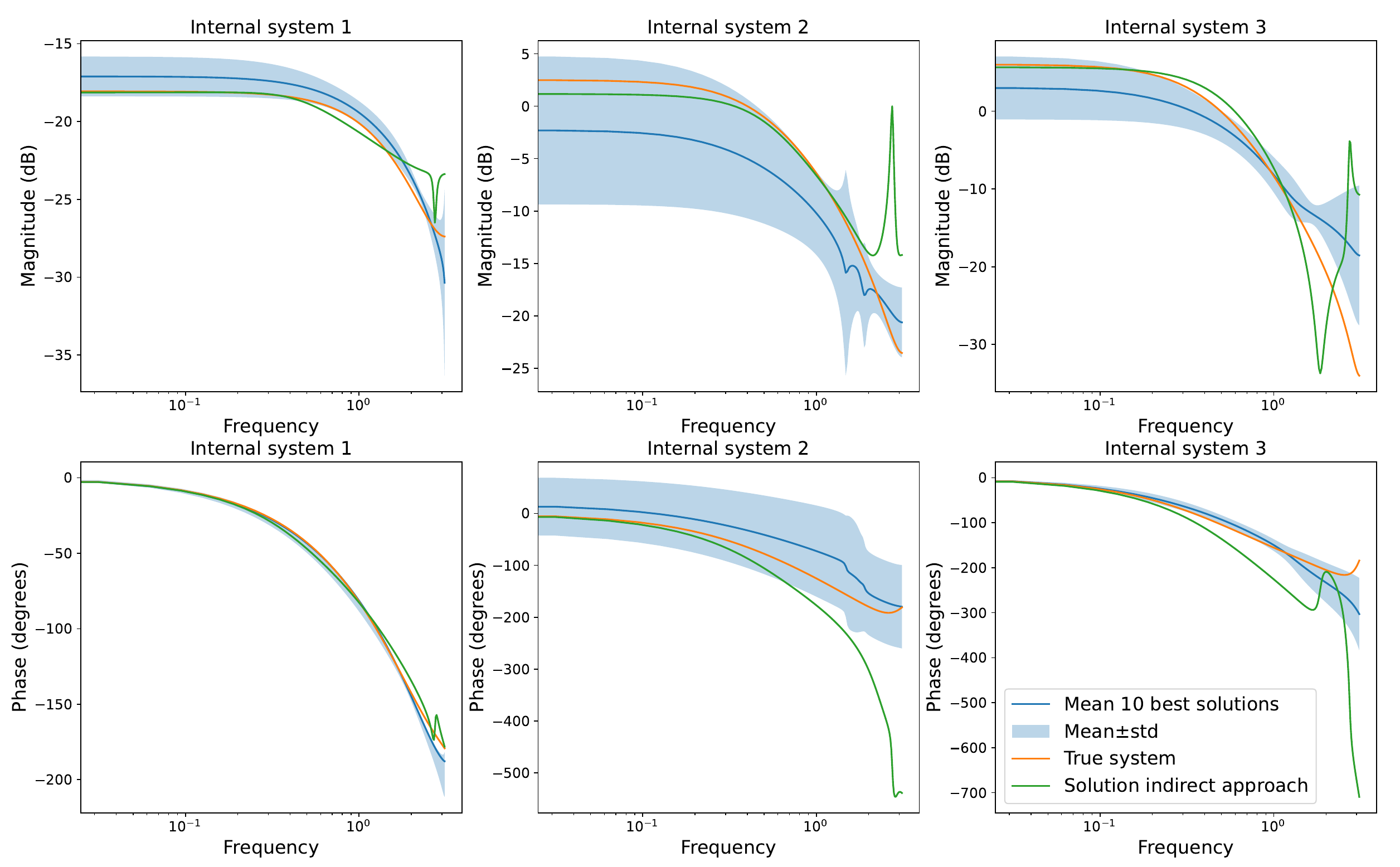}
\caption{Comparison of indirect and direct approach: magnitude and phase Bode plots for each internal system.} \label{bode_results}
\end{figure*}

As explained in Section~\ref{ssec:indirec_approach}, the indirect approach yields degree six polynomials for the subsystems, requiring a model reduction technique to allow a direct comparison between the parameters. Such a reduction may introduce additional errors in the estimate, so we instead compare the methods using a frequency domain analysis. Fig.~\ref{bode_results} presents such a comparison based on Bode plots for each subsystem, considering the ten best local solutions found using our approach, the solution found with the indirect approach, and the true system. 

From the Bode plots in Fig.~\ref{bode_results}, we can see that both approaches were able to obtain an estimate that somehow captures the dynamics of subsystems 1 and 3. However, when looking to the subsystem 2 the plots show a high variance when using the direct approach with different initial guesses. This highlights the necessity of implementing a good initialization strategy.

Next, we compare the model fit for the parameter estimates obtained using the direct and the indirect approach. 
Given a parameter estimate $\hat \theta$ and new realizations of the signals $(e^1,e^2,e^3)$ and the inputs $(r^1,r^2,r^3)$, the fit is defined as
\begin{align*}
    \text{fit}(\hat x) = 1 - \frac{\|\hat x - x^{\mathrm{ref}}\|_2}{\|\hat x - \mathbf{1}\bar x^{\mathrm{ref}} \|_2},
\end{align*}
where $\hat x$ is the signal of interest from the estimated model defined by $\hat \theta$, $x^{\mathrm{ref}}$ is the same signal from the true model, $\bar x^{\mathrm{ref}}$ denotes the mean of the true signal $x^{\mathrm{ref}}$, and $\mathbf{1}$ is a vector of all ones. 
With the estimate given by the indirect approach, the fit values for the observable and missing variables were $0.8934$ and $0.6312$, respectively, and with the best local solution from the direct approach, the corresponding fit values were $0.9008$ and $0.8344$, respectively. Analyzing these results we can check a clear advantage for the proposed direct approach, but when considering the other local solutions, we found that the fit associated with the missing variables $x_m$ rapidly decrease when going from the second best solution to the worst one, which also highlights the impact of and sensitivity to the choice of initial point.

As a simple initialization strategy, we tried using the reduced model obtained via the indirect approach as the initiation for the proposed direct approach. When comparing with the original fit value obtained with the indirect approach we had an improvement in the fit value for both the observed and the missing variables, which were $0.8973$ and $0.7253$, respectively. We repeated this experiment ten times with different realizations of both the error signals $(e^1,e^2,e^3)$ and the inputs $(r^1,r^2,r^3)$. Table I shows the fit of $x_o$ and $x_m$ for the high order models obtained by the indirect method and for the model obtained with the direct approach initialized based on a reduced model. The table also show the fit improvement, which is the difference between the direct approach and the indirect approach. We see that the two methods obtain a similar fit for $\hat x_o$, but we see a significant improvement in the fit for $x_m$. The mean improvement for $x_o$ and $x_m$ are $0.32\%$ and $12.99\%$, respectively. 

\begin{table}
  \label{tab:table_1}
  \hspace*{\dimexpr -\oddsidemargin}
    \caption{Fit values for the solutions from the indirect and the direct approach, and their respective increments.}
    \centering
  \begin{tabular}{|c|c|c|c|c|c|c|}
    \hline
    \multirow{2}{*}{Exp} &
      \multicolumn{3}{|c|}{$100 \times \text{Fit}(\hat x_o)$} &
      \multicolumn{3}{|c|}{$100 \times \text{Fit}(\hat x_m)$} \\
    & IND & DIR & IMPROV & IND & DIR & IMPROV \\
    \hline
    1 & 88.76 & 89.48 & 0.71 & 61.07 & 72.05 & 10.97 \\
    \hline
    2 & 88.86 & 89.78 & 0.92 & 38.14 & 68.38 & 30.24 \\
    \hline
    3 & 88.82 & 89.65 & 0.83 & 65.09 & 71.54 & 6.45 \\
    \hline
    4 & 88.87 & 89.41 & 0.54 & 66.31 & 76.68 & 10.37 \\
    \hline
    5 & 88.77 & 89.49 & 0.72 & 60.41 & 73.54 & 13.12 \\
    \hline
    6 & 89.24 & 89.75 & 0.50 & 63.08 & 76.60 & 13.52 \\
    \hline
    7 & 88.51 & 89.36 & 0.84 & 38.46 & 69.35 & 30.88 \\
    \hline
    8 & 88.94 & 89.43 & 0.49 & 61.78 & 67.24 & 5.46 \\
    \hline
    9 & 89.08 & 89.11 & 0.02 & 56.61 & 60.97 & 4.35 \\
    \hline
    10 & 88.87 & 86.42 & -2.44 & 57.71 & 62.23 & 4.51 \\
    \hline
  \end{tabular}
\end{table}

As indicated by the results, we observed that when only observing a few signals, our method often converges to a nearby local minimum at which the observable data $x_o$ is approximated well. However, if the starting point is not in a neighborhood of a global optimum, then we cannot ensure that estimate of $x_m$ will be accurate, but the estimate of $x_o$ is typically still good. We also observed that initialization based on another approach such as the indirect method can substantially improve the solution for the proposed direct approach.

\begin{figure*}[htbp]
    \centering
\includegraphics[width=\textwidth]{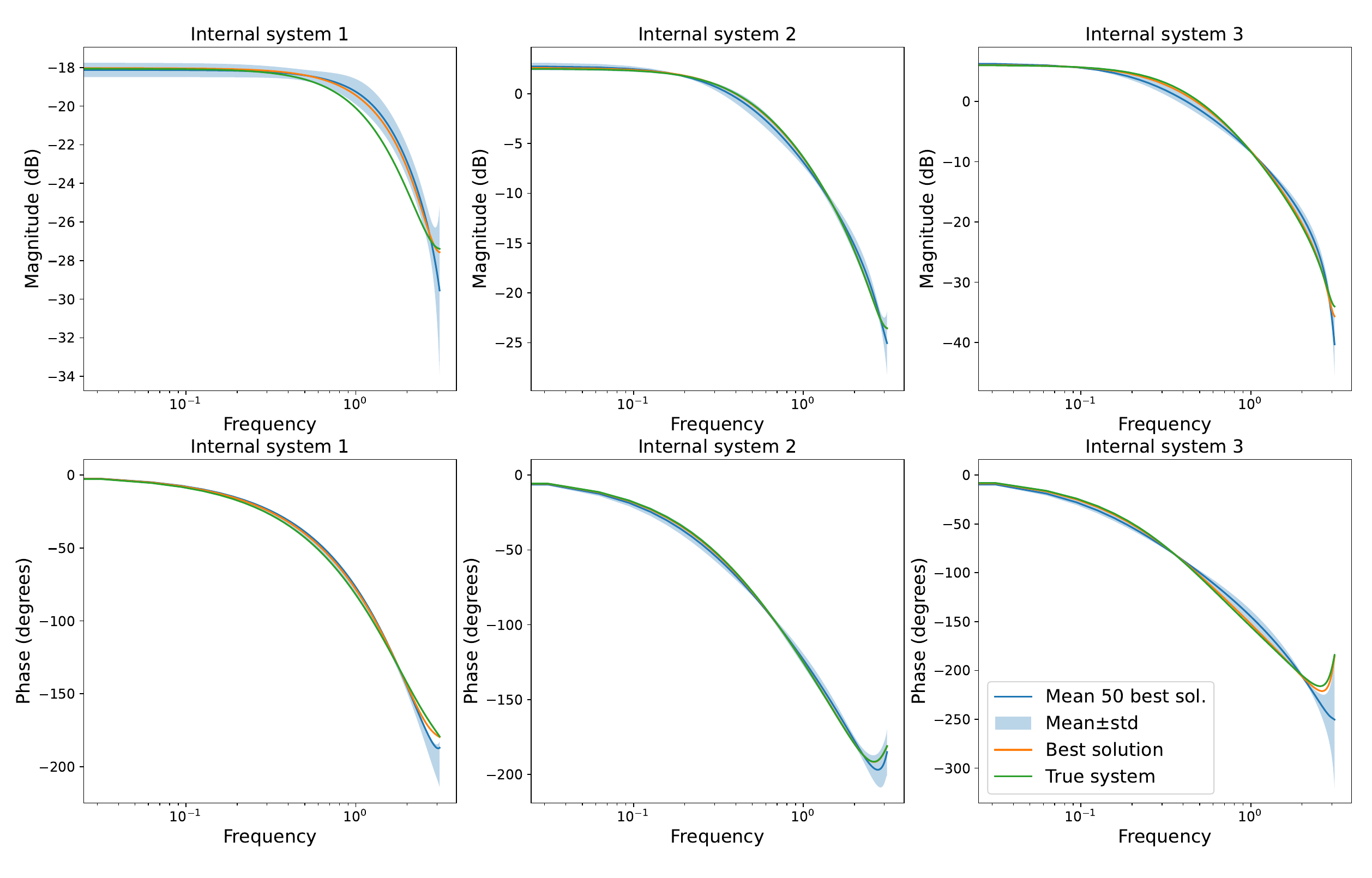}
\caption{Magnitude and phase Bode plots for each internal system, considering the 50 best solutions, when observing both $u^3$ and $u_1$ using the proposed direct approach.} \label{bode_results_2}
\end{figure*}

In contrast to the indirect approach where the knowledge of additional signals does not necessarily translate to better estimates, the proposed method can benefit from observing more signals, thereby improving the estimation accuracy and reducing the variance and sensitivity of the solutions for the different initial points. Fig.~\ref{bode_results_2} presents the Bode plots for each of the internal subsystems. The figure shows the true system and the estimate corresponding to the best local minimum along the mean and standard deviation for the 50 best local minima when observing both $u^1$ and $u^3$.

From the Bode plots in Fig.~\ref{bode_results_2}, we can see that observing $u^1$ together with $u^3$ leads to a considerable increase in the estimation accuracy while reducing the sensitivity to initialization.

Comparing the results in Figures~\ref{bode_results} and \ref{bode_results_2}, we see that with fewer observed signals, we are more likely to converge to a poor local minimum, and when we observe more variables, local optimization tends to end up in good local minima. This suggests that global optimization should be considered when only limited data are available.

%\addtolength{\textheight}{-3.7cm}   % This command serves to balance the column lengths

\section{Conclusion}\label{sec:conclusions}
We propose a new direct approach based on the ML estimation to perform system identification of dynamic networks with missing data. Dynamic networks generally lead to a singular pdf, and we show that when we know how the systems are interconnected, we are able to derive a nonsingular pdf by rewriting the problem using linear transformations, allowing us to perform the ML estimation.

The obtained results suggest that our approach is suitable for estimating the parameters of dynamic networks when combined with global optimization or a suitable initialization strategy. Using random initialization, we found that choosing the best local solution among many led to a precise prediction of the observable state and a good estimation of the missing ones.

As discussed in Section~\ref{ssec:indirec_approach}, we are not able to recover $C^i$ using the indirect approach. In contrast, our direct approach allows us to include this term, making it more suitable for ARMAX models. Another drawback of indirect approaches is that they can lead to unstable models for the subsystems even when the internal transfer functions are known to be stable. Moreover, the number of parameters that are required to describe the closed-loop system can be significantly larger. This either increases the variance of the estimator or requires the addition of constraints to the problem or a reparametrization of the cost function to avoid increasing the number of parameters.

The proposed direct approach can benefit from additional observed variables, while for the indirect approach, observing some of the signals does not add useful information. In the numerical example presented in Section~\ref{sec:num_exp}, knowing the sets $(u^3,u^1)$, $(u^3,y^3)$ or $(u^3,y^1)$ does not add more information when using the indirect approach than knowing just $u^3$. In contrast, using our method, each new observable variable would contribute to a better estimation. With our approach, the measurement of $(u^3,u^1)$ led to a considerable improvement in the fit values for both observed and missing variables, while decreasing the variance and sensitivity of the solutions for the different initial points.

When compared to prediction error based direct approaches, the proposed approach requires fewer observable variables to be able to estimate the system parameters, allowing the estimation of the network presented in Section~\ref{sec:num_exp} while observing only $u^3$, which would not be possible with prediction error based approaches \cite{comunitationVan_den_Hof}. 

The major limitation of our approach results from the nonconvexity of the cost function, requiring an initial set of parameters in the neighborhood of the global optimum to ensure the convergence towards it. Thus, a suitable initialization strategy is needed or, alternatively, the use of global optimization is called for. To investigate the issue, we explored two different approaches: the first approach was to consider 100 randomly generated initializations and use the best local solution with respect to the cost function value, and the second approach was to compute an initial guess from the estimate obtained using the indirect method. The results show that both approaches were able to converge to solutions for which the observable and the missing variables could be well estimated. 

In future work, we plan to apply the proposed approach to bigger dynamical network architectures, explore different initialization strategies and other optimization methods, e.g., trust-region methods or cheaper stochastic methods such as AdaHessian \cite{AdaHessian} or AdaSub \cite{AdaSub} that incorporate second-order information. We also plan to derive sufficiency conditions that characterize which signals need to be observed to obtain consistent estimates of each internal system in the dynamical networks when using the proposed maximum likelihood approach.

\IEEEtriggeratref{5}

\end{document}